\documentclass[a4paper]{article}

\usepackage{INTERSPEECH2021}
\usepackage{listings}
\usepackage{xcolor}
\usepackage{multirow}
\usepackage{diagbox}
\usepackage[hidelinks]{hyperref}
\usepackage{lipsum}

\colorlet{punct}{red!60!black}
\definecolor{background}{HTML}{EEEEEE}
\definecolor{delim}{RGB}{20,105,176}
\colorlet{numb}{magenta!60!black}

\lstdefinelanguage{json}{
    basicstyle=\normalfont\ttfamily,
    stepnumber=1,
    numbersep=6pt,
    showstringspaces=false,
    breaklines=true,
    frame=lines,
    literate=
     *{...}{{{\color{red}{...}}}}{3}
      {\{}{{{\color{delim}{\{}}}}{1}
      {\}}{{{\color{delim}{\}}}}}{1}
      {[}{{{\color{delim}{[}}}}{1}
      {]}{{{\color{delim}{]}}}}{1},
}

\title{GigaSpeech: An Evolving, Multi-domain ASR Corpus with \\10,000 Hours of Transcribed Audio}
\name{
  Guoguo Chen$^{*1,}\thanks{* Co-first authors, equal contribution.}$$^2$,
  Shuzhou Chai$^{*1,}$$^3$,
  Guanbo Wang$^{*1,}$$^{3,}$$^6$,
  Jiayu Du$^{*1}$,
  Wei-Qiang Zhang$^{*1,}$$^3$,\\
  Chao Weng$^{\dagger4}\thanks{$\dagger$ Authors listed in alphabetical order.}$,
  Dan Su$^{\dagger4}$,
  Daniel Povey$^{\dagger5}$,
  Jan Trmal$^{\dagger6}$,
  Junbo Zhang$^{\dagger5}$,
  Mingjie Jin$^{\dagger4}$,\\
  Sanjeev Khudanpur$^{\dagger6}$,
  Shinji Watanabe$^{\dagger6,}$$^7$,
  Shuaijiang Zhao$^{\dagger8}$,
  Wei Zou$^{\dagger8}$,
  Xiangang Li$^{\dagger8}$,\\
  Xuchen Yao$^{\dagger2}$,
  Yongqing Wang$^{\dagger5}$,
  Yujun Wang$^{\dagger5}$,
  Zhao You$^{\dagger4}$,
  Zhiyong Yan$^{\dagger5}$,
}
\address{
  $^1$SpeechColab
  $^2$Seasalt AI Inc
  $^3$Dept EE, Tsinghua University
  $^4$Tencent AI Lab
  $^5$Xiaomi Corporation\\
  $^6$CLSP \& HLTCOE, The Johns Hopkins University
  $^7$Carnegie Mellon University
  $^8$KE Holdings Inc}
\email{gigaspeech@speechcolab.org}

\begin{document}
\eightpt
\maketitle
\begin{abstract}
This paper introduces GigaSpeech, an evolving, multi-domain English speech recognition
  corpus with 10,000 hours of high quality labeled audio suitable for supervised training,
  and 40,000 hours of total audio suitable for semi-supervised and unsupervised training.
  Around 40,000 hours of transcribed audio is first collected from audiobooks, podcasts and
  YouTube, covering both read and spontaneous speaking styles, and a variety of
  topics, such as arts, science, sports, etc. A new forced alignment and segmentation pipeline is proposed
  to create sentence segments suitable for speech recognition training, and to
  filter out segments with low-quality transcription. For system training, GigaSpeech provides five subsets of different sizes, 10h, 250h, 1000h, 2500h, and 10000h. For our 10,000-hour
  {\it XL} training subset, we cap the word error rate at 4\% during the
  filtering/validation stage, and for all our other smaller training subsets,
  we cap it at 0\%. The {\it DEV} and {\it TEST} evaluation sets, on the other
  hand, are re-processed by professional human transcribers to ensure high
  transcription quality. Baseline systems are provided for popular speech
  recognition toolkits, namely Athena, ESPnet, Kaldi and Pika.
\end{abstract}


\section{Introduction}
Thanks to the rapid development of the neural network models, automatic speech recognition (ASR)
has made tremendous progress in the past decade. Various system architectures,
from hybrid \cite{dahl2011context} to end-to-end \cite{graves2014towards}, are
proposed, and state-of-the-art results on standard benchmarks are being
frequently updated.

The mainstream speech recognition corpora, on the other
hand, have not changed much in decades. To take the English speech recognition
task as an example, the Wall Street Journal corpus, which consists of 80 hours
of narrated news articles \cite{paul1992design}, is almost 20 years old, and has
a word error rate (WER) of 2.32\% on its {\it eval92} benchmark \cite{Povey2016Purely}.
The Switchboard and Fisher corpus, which consists of 262 and 1,698 hours of
telephone conversational speech, is also around 20 years old, and has a WER of
5.5\% on the Switchboard portion of the {\it Hub5'00} benchmark \cite{saon2017english}.
Even LibriSpeech \cite{panayotov2015librispeech}, one of the most popular corpora
for speech recognition tasks, is more than 5 years old, and has a WER of
1.9\% on its {\it test\_clean} benchmark \cite{gulati2020conformer}. It consists of
1,000 hours of read English speech. Due to the fast development of speech
recognition techniques, ASR performance on those data sets appears to have saturated,
making it difficult to track further improvements from new techniques.

There is some progress on creating better corpora/benchmarks for English speech
recognition, from both academia and industry. TED-LIUM \cite{hernandez2018ted} is a
series of corpora created by the Ubiqus company and the University of Le Mans.
It consists of 452 hours of audio from TED talks in its latest release
TED-LIUM 3. The corpora size, however, is less than 1,000 hours, making it not
suitable for algorithms which demand a large amount of data. People's Speech \cite{peoplespeech}
released by ML Commons consists of 87,000 hours of audio, covering 59
different languages. It's source, however, is mostly audiobook, lacking
crucial acoustic diversity. Another work is SPGISpeech \cite{oneill21},
a corpus released by Kensho Technologies. It consists of 5,000 hours of transcribed audio
from earnings calls transcribed by S\&P Global, Inc. The corpus by its nature
gives an emphasis to the business domain.

We release a complementary English speech recognition corpus named GigaSpeech,
an evolving, multi-domain ASR corpus with 10,000 hours of transcribed Audio.
The initial release of GigaSpeech is complementary to the existing corpora in
the following ways:
\begin{itemize}
  \item Extensible, the metadata is designed in a way that it can be easily reused
    for other tasks, such as speaker identification.
  \item Large scale, 10,000 hours of transcribed speech.
  \item Multi-source, covers audiobook, podcast and YouTube.
  \item Multi-style, covers both read and spontaneous speech.
  \item Multi-topic, covers a variety of topics, such as arts, science, sports, etc.
  \item Original/Normalized transcription pairs, suitable for training end-to-end
    systems with post-processing (punctuation, case/date/time normalization, etc) included.
\end{itemize}

We make two contributions in this work. First, we release an evolving,
multi-domain speech recognition corpus with 10,000 hours of labeled audio.
Second we provide a scalable, reliable pipeline for generating speech
recognition corpora.

The rest of the paper is organized as follows. Section 2 introduces the
GigaSpeech corpus, and Section 3 presents the full pipeline to create the
GigaSpeech corpus. We describe the speech recognition baseline systems for various
toolkits, and provide experiment setup and results in Section 4.
Finally, acknowledgements are given in Section 5.

\section{GigaSpeech Corpus}
This section explains the structure of the GigaSpeech corpus, including
metadata, data partition, audio format, etc. Instructions and scripts for downloading GigaSpeech can be found on GigaSpeech's GitHub repository\footnote{\url{https://github.com/SpeechColab/GigaSpeech}\label{foot:gigaspeech_repo}}.

\subsection{Metadata}
\begin{figure}[t]
  \centering
  \includegraphics[width=\linewidth]{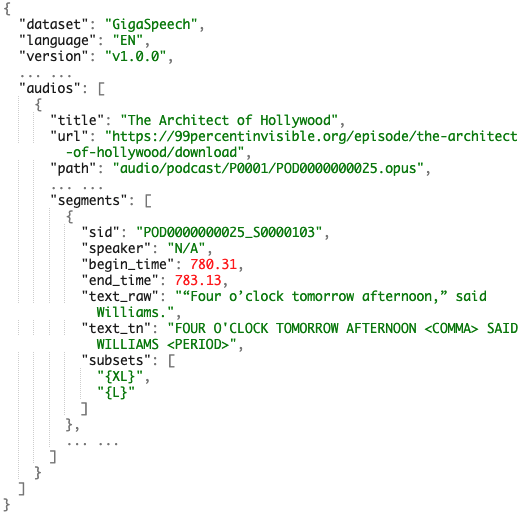}
  \vspace{-5ex}
  \caption{A snip of the metadata file GigaSpeech.json}
  \label{fig:metadata}
\end{figure}

We save all the metadata information to a single JSON file named GigaSpeech.json.
Figure \ref{fig:metadata} shows a snip of this file. For better presentation of
this paper, we skip a lot of non-critical entries in the snip, such as
``format", ``md5", ``source", etc. 

To use the corpus, users are expected to extract the relevant information from
GigaSpeech.json. For example, for the speech recognition task, one should first
follow the ``audios" entry, and work out a list of audio files. One can then
follow the ``url" entry to download the original audio file, or ``path" if
preprocessed audio files have been downloaded to the disk. After that, for each
audio file, one can follow the ``segments" entry, and work out the trainable
audio segments, as well as their corresponding transcripts. Of course, we also
have various supplementary entries, such as ``subsets", ``md5", which will also
be helpful for your task.

The metadata file GigaSpeech.json is version controlled, and is supposed to get
updated over the time. In future releases, we plan to add speaker information
to the metadata file, so that it will be suitable for speaker
identification/verification tasks. We also plan to add more data from different
sources to increase the diversity.

\subsection{Training Subsets}

\begin{table}[t]
  \caption{GigaSpeech training subsets}
  \vspace{-2ex}
  \label{tab:train_subsets}
  \centering
  \begin{tabular}{c|rrr|r}
    \toprule
    \textbf{Subset} & \textbf{Audiobook} & \textbf{Podcast} & \textbf{YouTube} & \textbf{Total} \\
    \midrule
    {\it XL}         & 2,655h             & 3,499h           & 3,846h           & 10,000h \\
    {\it L}          & 650h               & 875h             & 975h             & 2,500h \\
    {\it M}          & 260h               & 350h             & 390h             & 1,000h \\
    {\it S}          & 65h                & 87.5h            & 97.5h            & 250h \\
    {\it XS}         & 2.6h               & 3.5h             & 3.9h             & 10h \\
    \bottomrule
  \end{tabular}
  \vspace{-2ex}
\end{table}

We provide 5 training subsets in GigaSpeech, namely {\it XS}, {\it S}, {\it M},
{\it L} and {\it XL}, listed here in order of increasing audio hours.
Table \ref{tab:train_subsets} shows a detailed breakdown of the 5 GigaSpeech
training subsets.

\subsection{Evaluation Sets}

\begin{table}[t]
  \caption{GigaSpeech evaluation sets}
  \vspace{-2ex}
  \label{tab:eval_sets}
  \centering
  \begin{tabular}{c|rr|r}
    \toprule
    \textbf{Sets} & \textbf{Podcast} & \textbf{YouTube} & \textbf{Total} \\
    \midrule
    {\it DEV}        & 6.3h          & 6.2h             & 12.5h \\
    {\it TEST}       & 16.1h         & 24.2h            & 40.3h \\
    \bottomrule
  \end{tabular}
\end{table}

We provide 2 evaluation sets in GigaSpeech: a {\it DEV} set for development and
tuning, which consists of 12.5 hours of audio, and a {\it TEST} set for
final evaluation, which consists of 40.3 hours of audio.

A breakdown of our evaluation sets is illustrated in Table \ref{tab:eval_sets}.
Note that our evaluation sets do not have a coverage for the audiobooks.
We make sure that audio files from the LibriSpeech \cite{panayotov2015librispeech} evaluation
sets ({\it dev-clean}, {\it dev-other}, {\it test-clean} and {\it test-other})
are not presented in our corpus, therefore, the LibriSpeech evaluation sets
can be used as our evaluation sets as well.

\subsection{Audio Format}

\begin{table}[t]
  \caption{Impact of Opus audio compression (WER in \%)}
  \vspace{-2ex}
  \label{tab:eval_opus}
  \centering
  \begin{tabular}{c|c|rr|rr}
    \toprule
    \multicolumn{2}{c|}{\multirow{2}{*}{\diagbox{Train}{Eval}}} & \multicolumn{2}{c|}{{\it DEV}}     & \multicolumn{2}{c}{{\it TEST}}\\
    \cline{3-6}
    \multicolumn{2}{c|}{}                                       & Opus       & Wav                  & Opus          & Wav \\
    \midrule
    \multirow{2}{*}{{\it M}} & Opus                           & 19.0      & 18.7                & 18.5         & 18.3\\
     & Wav                            & 18.8      & 18.5                & 18.3         & 18.2\\
    \bottomrule
  \end{tabular}
  \vspace{-2ex}
\end{table}

To reduce the file size of the GigaSpeech corpus, we compress the original
audio using the Opus audio codec. Original audio files are first converted to
16k sampling rate, single channel and 16-bit signed-integer format. Opus
compression is then applied to achieve an output bit rate of 32 kpbs, which
results in a compression ratio of 8.

Table \ref{tab:eval_opus} shows the impact of Opus audio compression in terms
of WER (\%). Kaldi systems (see Section \ref{sec:kaldi_baseline}, but without recurrent neural network language model rescoring) are built for
our {\it M} (1000h) training subset, with or without Opus compression. These two systems
are then used to decode the {\it DEV} and {\it TEST} evaluation sets, with or
without Opus compression. From Table \ref{tab:eval_opus}, it is clear that
compressing the training data with the Opus codec at 32 kpbs output bit rate has very small impact on the {\it DEV} and {\it TEST} set (0.1 - 0.2\% WER degradation).

\section{GigaSpeech Creation Pipeline}
This section presents the detailed pipeline for creating the GigaSpeech corpus,
which can be applied to other data generation tasks as well.

\subsection{Stage 1: Audio Collection}
We start the task by manually defining the categories that we are interested in.
We selected 24 categories in total, namely
{\it Arts}, {\it Business}, {\it Education}, {\it Autos and Vehicles},
{\it Comedy}, {\it Crime}, {\it Entertainment}, {\it Film and Animation},
{\it Gaming}, {\it Health and Fitness}, {\it History}, {\it Howto and Style},
{\it Kids and Family}, {\it Leisure}, {\it Music}, {\it News and Politics},
{\it Nonprofits and Activism}, {\it People and Blogs}, {\it Pets and Animals}, {\it Religion and Spirituality},
{\it Science and Technology}, {\it Society and Culture}, {\it Sports}, {\it Travel and Events}.

For podcasts, we follow the above categories, and select episodes that come with
manual transcriptions. For YouTube, we use the above categories as seed
keywords, and select videos with human-generated  closed captions. For audiobooks, we
do not enforce those categories.

Once we have the list of audio files, we create tools and download all audio files
with their corresponding transcripts.

\subsection{Stage 2: Text Normalization}
The audio transcripts we download from various sources are created by
different transcribers with diversified transcription standards and styles,
therefore it is necessary to apply text normalization to the original
transcripts. We perform standard text normalization, including case normalization,
special symbol removal, number to word rewriting, date/time rewriting, etc.


For audiobooks and podcasts, transcripts are usually at the episodes or chapter/book
level. For speech recognition, however, smaller segments less than 20 seconds are
needed for training. The next step is to segment the long audio file into smaller
segments. For YouTube, closed captions are provided at the sentence
level, but unfortunately we find that the timestamps of closed captions are not
reliable for segmentation. As a result, we decide to splice the closed captions
all together, and perform the same segmentation as audiobooks and podcasts.

\subsection{Stage 3: Forced Alignment}
Our aligner is implemented with Kaldi \cite{povey2011kaldi}, and the alignment
procedure follows the work here \cite{manohar2017jhu}, which adopts a divide and
conquer strategy to tackle the alignment problem. First, both audio and transcript
are uniformly chunked into smaller pieces. Second, audio segments are decoded with
a biased language model (LM), and hypotheses with timestamps are generated. Third, each hypothesis segment is matched to one transcript segment via TF-IDF similarity. For each matched pair, 
hypothesis is further aligned with the transcript segment using
the Smith-Waterman algorithm \cite{pearson1991searching}. Finally, through this alignment, timestamps are attached to the transcript segments,
and eventually to the whole transcripts by stitching the independently aligned segments together. Note that we modify the Smith-Waterman algorithm to handle silence and punctuation, and this is essential to enable the sentence-based segmentation in the next section.

To achieve better alignment performance, we first align and segment the
downloaded audio with a close-domain acoustic model. We then train an in-domain
acoustic model with the audio segments created (around 3,000 hours). This model
is used to align the whole corpus.

\subsection{Stage 4: Audio Segmentation}
We work out the audio segments from the alignment information above. Several
rules are applied during the segmentation process:
\begin{itemize}
  \item Split allowed at silence that is longer than 1 second.
  \item Split allowed at punctuation (``,", ``.", ``!" or ``?") that is
    longer than 0.2 seconds.
  \item Segments with alignment WER $\geq$ 75\% re removed.
  \item Segments with length $\geq$ 20 seconds are removed.
  \item Silence at segment boundaries are truncated to 0.15 seconds.
\end{itemize}

It is worth pointing out that we keep 4 types of punctuation, namely
comma, period, question mark and exclamation mark, so that split can happen at sentence boundaries (second rule above). We map them to special
words ``$<$COMMA$>$", ``$<$PERIOD$>$", ``$<$QUESTIONMARK$>$" and
``$<$EXCLAMATIONMARK$>$" respectively. Besides, this also allows us to build end-to-end
speech recognition systems that includes punctuation tagging, and end
point detection.

\subsection{Stage 5: Segment Validation}
The segmentation stage generates a list of candidate segments, but potentially
with high transcription error rate. We therefore apply segment validation to
filter out bad segments.

\subsubsection{Forced Alignment Graph}
\begin{figure}[t]
  \centering
  \includegraphics[width=\linewidth]{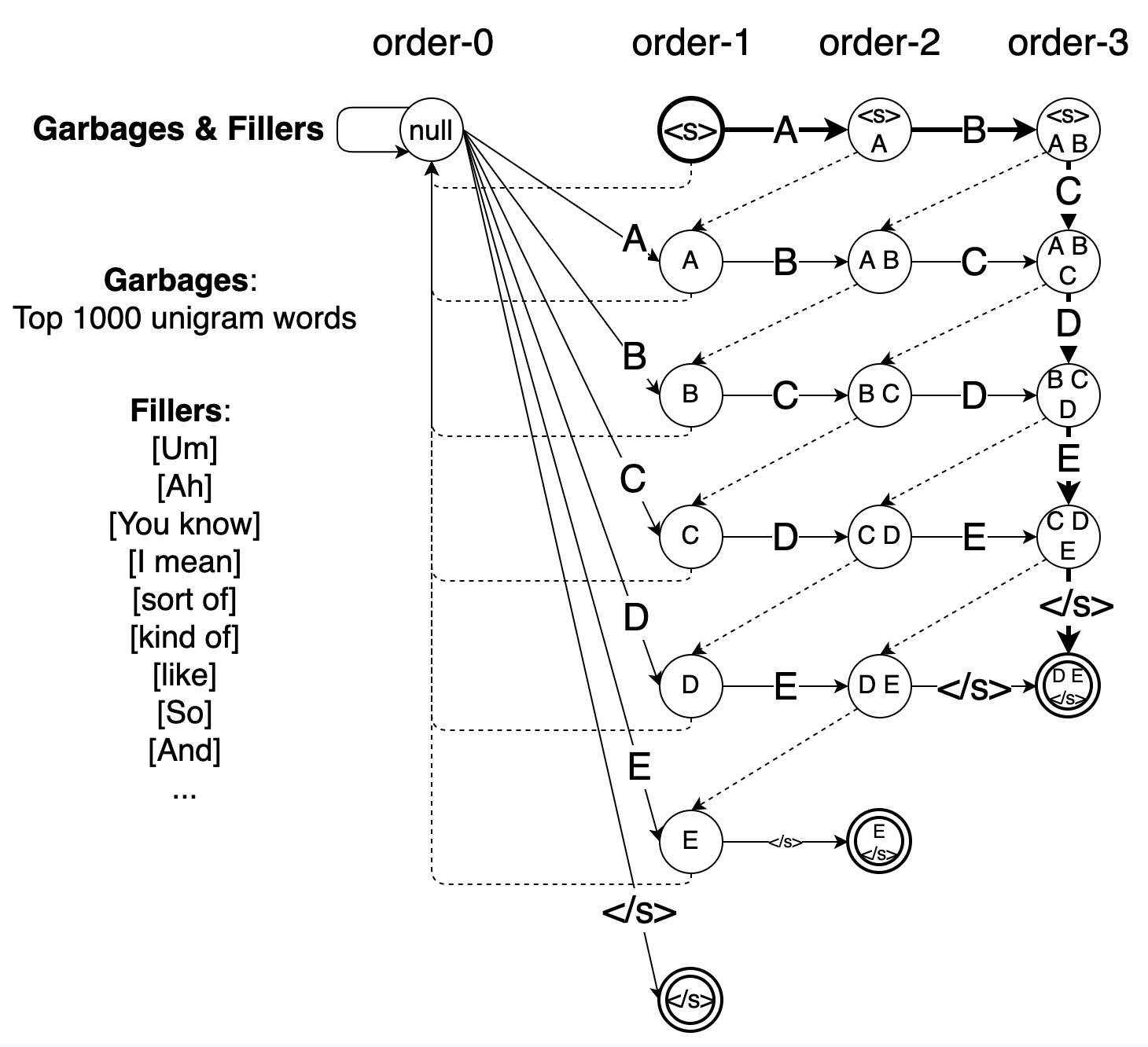}
  \vspace{-5ex}
  \caption{A forced alignment graph for the sentence ``$<$s$>$ A B C D E $<$/s$>$" (4-gram)}
  \label{fig:alignment_graph}
  \vspace{-2ex}
\end{figure}

To better detect transcription errors made by human transcribers, we propose a
variation of alignment graph in n-gram framework, as shown in Figure \ref{fig:alignment_graph}.
The bold arrow path represents a typical LM-free forced alignment graph.
Each state on the forced alignment path has a dotted ``leaky" arc (with weight) that allows
the token to leak out the forced alignment path, from higher order n-gram states
down to lower order n-gram states, until reaching the null state. A garbage word
loop (containing top 1,000 uni-gram words) is added around the null state to consume
additional acoustic frames. Besides, there are extra states and arcs that allow the token
to return to the forced alignment path. In general, this alignment graph allows
the decoder to perform insertion/deletion/substitution to the reference. This essentially brings more flexibility to the forced alignment stage, making it possible to capture the discrepancy between the audio and the corresponding transcript.

\subsubsection{Validation Decoding Pass}
\begin{figure*}[t]
  \centering
  \vspace{-2ex}
  \includegraphics[width=\linewidth]{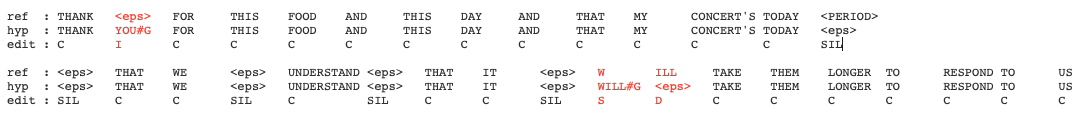}
  \vspace{-4ex}
  \caption{Examples of transcription errors detected by forced alignment}
  \vspace{-2ex}
  \label{fig:alignment_rewriting}
\end{figure*}

During the validation decoding pass, we detect transcription errors and filter
out segments with high error rate. Figure \ref{fig:alignment_rewriting} gives
two examples of how errors are being detected. In the first example, the
transcriber misses a word ``YOU" in the transcript, which is caught by our
decoder. In the second example, the transcriber writes a typo which is also
successfully detected.

For the podcast and YouTube portion of our {\it XL} training subset, we cap the
maximum WER at 4\%, and throw away all segments with higher WER. For the
audiobook portion of the {\it XL} training subset, as well as all other smaller
subsets, we cap the maximum WER at 0\%, meaning we don't allow any transcription
errors.

\subsubsection{Reference Rewriting}
Investigation into the validated segments reveals three common types of transcriber errors:
\begin{itemize}
  \item Fillers ignored, such as AH, UH, UM, ER, ERR, YOU KNOW, I MEAN, SORT OF, etc.
  \item Conjunctions ignored, such as  AND, OR, BUT, etc.
  \item Disfluency removed, such as ``It's it's it's a great thing!".
\end{itemize}

Discarding these segments will fundamentally limit the diversity of the corpus. To fix those common errors, we add a filler loop (see Figure \ref{fig:alignment_graph}) to the forced alignment graph, which contains the above common fillers and
conjunctions. Besides, we also employ a disfluency detector. The filler loop and the disfluency detector may modify the reference (reference rewriting), and if that happens, those words will be counted as correct. We only apply reference rewriting to our {\it XL} subset (10,000h).

\subsection{Stage 6: Evaluation}
\begin{figure}[t]
  \centering
  \includegraphics[width=\linewidth]{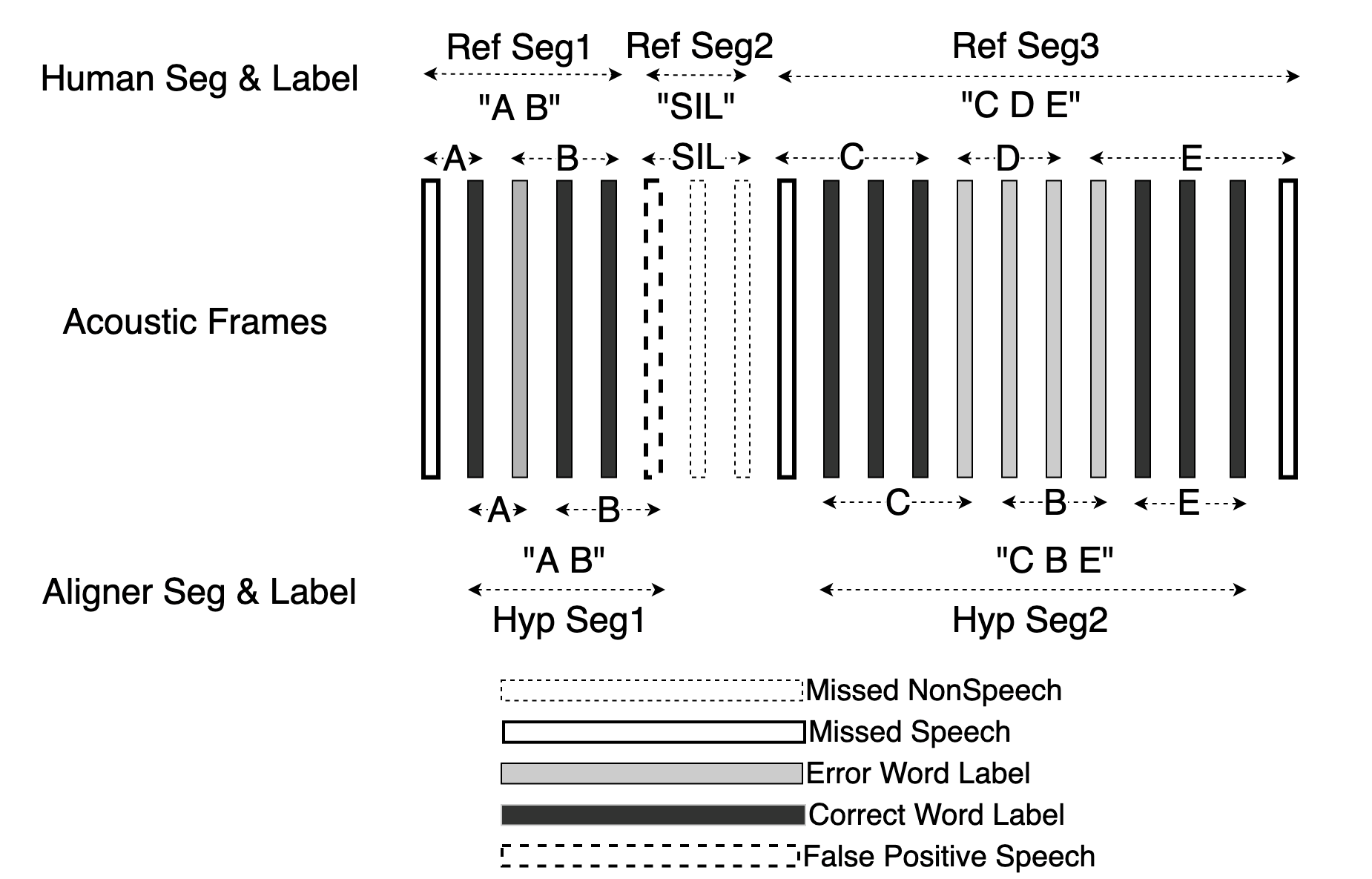}
  \vspace{-4ex}
  \caption{Frame level segmentation accuracy}
  \vspace{-1ex}
  \label{fig:frame_accuracy}
\end{figure}

Since our evaluation sets are manually processed by professional human transcribers,
we take that as the ground truth, and use it to compute the frame level segmentation
precision and recall, see Figure \ref{fig:frame_accuracy}. Here recall tells us
how many frames can be retrieved by our segmentation and validation pipeline, and
precision tells us how many of these retrieved frames are correctly labeled (consistent
with the human labels).

\begin{figure}[t]
  \centering
  \includegraphics[width=\linewidth]{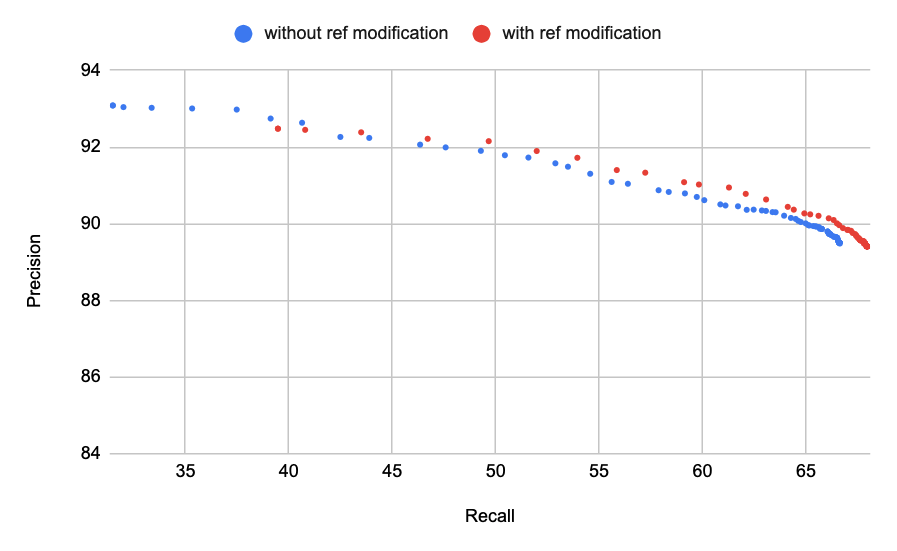}
  \vspace{-6ex}
  \caption{Frame level precision-recall curve for the segmentation and validation pipeline}
  \vspace{-4ex}
  \label{fig:frame_roc}
\end{figure}

Figure \ref{fig:frame_roc} illustrates the precision-recall curve on the podcast
portion of the {\it DEV} evaluation set. For the {\it XL} training subset, we
select a working point that gives us 10,000 hours of validated audio data, while
keeping the maximum WER under 4\%. And for all our other training subsets, we
select the working point that keeps the maximum WER at 0\% (the left most
working point on Figure \ref{fig:frame_roc}).

\section{Experiments}
This section describes the baseline systems and experimental results for four popular speech recognition
toolkits, namely Athena, ESPnet \cite{Watanabe2018ESPnet},
Kaldi \cite{povey2011kaldi} and Pika \cite{Weng2020}.

\subsection[Athena Baseline System]{Athena Baseline System\footnote{\url{https://github.com/athena-team/athena/tree/master/examples/asr/gigaspeech}}\label{foot:athena}}
The Athena baseline implements an encoder-decoder based transformer model, which
is similar to \cite{karita2019comparative}, with parameters
$e=12$, $d=6$, $d_{\rm model}=1024$, $d_{\rm ff}=2048$ and $d_{\rm head}=8$.

During the training, the output sequence of the encoder is also used for
connectionist temporal classification (CTC) for joint training to enforce
monotonic alignment between speech and label sequences. We use the Adam
optimizer and varied learning rate with a warmup schedule (warmup steps = 8000).
The model is trained with a total batch size of 128 for 5 epochs.

During the decoding, beam search with a beam size of 8 is used, which combines
the scores of the decoder, the CTC with weight 0.5 and the LM with weight 0.2.
The LM is a RNN based model with two long short-term memory (LSTM) layers, each
with 1024 nodes.

\subsection[ESPnet Baseline System]{ESPnet Baseline System\footnote{\url{https://github.com/espnet/espnet/tree/master/egs2/gigaspeech/asr1}}\label{foot:espnet}}
The ESPnet baseline uses conformer (Convolution-Augmented Transformer) \cite{gulati2020conformer}, which is a recently proposed architecture combining the local sensitivies of convolutional neural networks with the long-range interactions of transformers \cite{vaswani2017attention}.
We use the implementation provided in the ESPnet toolkit \cite{guo2020recent}.
We use a set of 5k BPE tokens generated by the SentencePiece tokenizer \cite{kudo2018sentencepiece}.
The model has a 12 conformer blocks with an output dimension of 512 and a kernel size of 31 in the encoder, and 6 transformer blocks in the decoder.
Both encoder and decoder have 8 attention heads with 2048 feed-forward unit dimension.
Conformer was trained using four 24Gb memory Titan RTX GPUs. The max trainable epoch is 20. Mini-batch size is 35 million acoustic feature bins. Adam optimizer with no weight decay was used. Noam learning rate scheduler was set to 25k warmup steps with a learning rate of 0.0015. SpecAug used 2 frequency masks and 5 time masks. The last 10 best checkpoints were averaged as the final model.

\subsection[Kaldi Baseline System]{Kaldi Baseline System\footnote{\url{https://github.com/kaldi-asr/kaldi/tree/master/egs/gigaspeech}}\label{foot:kaldi}} \label{sec:kaldi_baseline}
The Kaldi baseline implements a typical chain model. First, a GMM-HMM model is trained to obtain the alignments with no data cleaning. Second, volume and speed augmentation techniques are applied. I-vectors are then extracted and pasted to the basic acoustic features, as most Kaldi's recipes do. Finally, a neural network is trained with both cross-entropy and LF-MMI criteria. Our neural network stacks 6 convolutional neural network (CNN) layers, 10 TDNN-F layers, 1 attention-relu-renorm-layer, 1 TDNN-F layer, 1 fast-lstmp-layer, 1 TDNN-F layer and finally 1 more fast-lstmp-layer. During the decoding stage, a 4-gram LM is first used for decoding, followed by a Recurrent Neural Network Language Model (RNNLM) rescoring pass.

\subsection[Pika Baseline System]{Pika Baseline System\footnote{\url{https://github.com/tencent-ailab/pika/tree/main/egs/gigaspeech}}\label{foot:pika}}
The Pika baseline adopts a convolution and transformer based architecture \cite{Weng2020} for the encoder of our RNN-T system. Five-layer transformer is used for the decoder and the hidden dimension of each layer is 512. We apply on-the-fly speed and volume perturbation during training where speed rates are set to 0.9/1.0/1.1/1.2 and the volume range is from -55dB to -10dB. For input features, we use 80 dimensional log Fbanks. The targets of our RNN-T system are a set of English wordpieces plus blank symbol which lead to an output dimension of 5000. The number of total parameters in the RNN-T is about 87M. MBR training and two extra two-layer transformer based forward/backward rescorers are also adopted. All training is conducted on 16 V100 GPUs. Our distributed training strategy is based on block-wise model-update filtering (BMUF) with a Nesterov momentum scheme but with different learning rate scheduling \cite{Weng2020} where both initial and final learning rates are set before training and number of training
epochs are fixed (therefore there is no early stop and no
development/validation set is used). One single sweep of the GigaSpeech \textit{XL} training subset takes about 5hrs. For decoding, we set beam-size to 8 and the temperature of softmax to 1.25.

\subsection{Experimental Results}
\begin{table}[t]
  \caption{GigaSpeech baselines for the \textit{XL} training subset (WER in \%)}
  \vspace{-2ex}
  \label{tab:toolkit_baseline}
  \centering
  \begin{tabular}{c|c|rr}
    \toprule
    \textbf{Toolkit}                      & \textbf{Model}               & \textbf{\textit{DEV}}    & \textbf{\textit{TEST}} \\
    \midrule
    Athena                                 & Transformer-AED + RNNLM      & 13.60                & 12.70 \\
    ESPnet                                 & Conformer/Transformer-AED    & 10.90                & 10.80 \\
    Kaldi                                  & Chain + RNNLM                & 14.78                & 14.84 \\
    Pika                                   & RNN-T                        & 12.30                & 12.30 \\
    \bottomrule
  \end{tabular}
\end{table}

\begin{table}[t]
  \caption{Kaldi baselines for GigaSpeech training subsets (WER in \%)}
  \vspace{-2ex}
  \label{tab:kaldi_baseline}
  \centering
  \begin{tabular}{c|rr}
    \toprule
    \textbf{Subset}                        & \textbf{\textit{DEV}}    & \textbf{\textit{TEST}} \\
    \midrule
    \textit{XL}                            & 14.78                    & 14.84 \\
    \textit{L}                             & 16.60                    & 16.28 \\
    \textit{M}                             & 17.96                    & 17.53 \\
    \textit{S}                             & 22.59                    & 22.14 \\
    \textit{XS}                            & N/A                      & N/A \\
    \bottomrule
  \end{tabular}
  \vspace{-4ex}
\end{table}

Table \ref{tab:toolkit_baseline} demonstrates baseline results for Athena, ESPnet, Kaldi and Pika. Results listed here are purely for the purpose of providing baseline systems for each toolkit. They do not reflect the state-of-the-art performance of each toolkit, and cannot be used to compare the performance across toolkits.

Table \ref{tab:kaldi_baseline} illustrates the Kaldi baseline results for 4 training subsets. Generally speaking, as the training subset gets bigger, the performance goes up. Our smallest {\it XS} training subset (10h) is designed for system building and debugging only, and is not expected to give strong performance.

\section{Acknowledgements}
The authors would like to thank Xingyu Na for his various suggestions on the
Kaldi baseline system. The authors also would like to thank Speechocean for
transcribing the GigaSpeech evaluation sets.

\bibliographystyle{IEEEtran}

\bibliography{refs}

\begin{thebibliography}{10}
\providecommand{\url}[1]{#1}
\csname url@samestyle\endcsname
\providecommand{\newblock}{\relax}
\providecommand{\bibinfo}[2]{#2}
\providecommand{\BIBentrySTDinterwordspacing}{\spaceskip=0pt\relax}
\providecommand{\BIBentryALTinterwordstretchfactor}{4}
\providecommand{\BIBentryALTinterwordspacing}{\spaceskip=\fontdimen2\font plus
\BIBentryALTinterwordstretchfactor\fontdimen3\font minus
  \fontdimen4\font\relax}
\providecommand{\BIBforeignlanguage}[2]{{%
\expandafter\ifx\csname l@#1\endcsname\relax
\typeout{** WARNING: IEEEtran.bst: No hyphenation pattern has been}%
\typeout{** loaded for the language `#1'. Using the pattern for}%
\typeout{** the default language instead.}%
\else
\language=\csname l@#1\endcsname
\fi
#2}}
\providecommand{\BIBdecl}{\relax}
\BIBdecl

\bibitem{dahl2011context}
G.~E. Dahl, D.~Yu, L.~Deng, and A.~Acero, ``Context-dependent pre-trained deep
  neural networks for large-vocabulary speech recognition,'' \emph{IEEE
  Transactions on Audio, Speech, and Language Processing}, vol.~20, no.~1, pp.
  30--42, 2011.

\bibitem{graves2014towards}
A.~Graves and N.~Jaitly, ``Towards end-to-end speech recognition with recurrent
  neural networks,'' in \emph{Proc. International Conference on Machine
  Learning (ICML)}.\hskip 1em plus 0.5em minus 0.4em\relax PMLR, 2014, pp.
  1764--1772.

\bibitem{paul1992design}
D.~B. Paul and J.~Baker, ``The design for the wall street journal-based csr
  corpus,'' in \emph{Proc. International Conference on Spoken Language
  Processing (ICSLP)}.\hskip 1em plus 0.5em minus 0.4em\relax ISCA, 1992.

\bibitem{Povey2016Purely}
D.~Povey, V.~Peddinti, D.~Galvez, P.~Ghahremani, V.~Manohar, X.~Na, Y.~Wang,
  and S.~Khudanpur, ``{Purely Sequence-Trained Neural Networks for ASR Based on
  Lattice-Free MMI},'' in \emph{Proc. Interspeech 2016}.\hskip 1em plus 0.5em
  minus 0.4em\relax ISCA, 2016, pp. 2751--2755.

\bibitem{saon2017english}
G.~Saon, G.~Kurata, T.~Sercu, K.~Audhkhasi, S.~Thomas, D.~Dimitriadis, X.~Cui,
  B.~Ramabhadran, M.~Picheny, L.-L. Lim \emph{et~al.}, ``English conversational
  telephone speech recognition by humans and machines,'' \emph{arXiv preprint
  arXiv:1703.02136}, 2017.

\bibitem{panayotov2015librispeech}
V.~Panayotov, G.~Chen, D.~Povey, and S.~Khudanpur, ``Librispeech: an asr corpus
  based on public domain audio books,'' in \emph{Proc. 2015 IEEE International
  Conference on Acoustics, Speech and Signal Processing (ICASSP)}.\hskip 1em
  plus 0.5em minus 0.4em\relax IEEE, 2015, pp. 5206--5210.

\bibitem{gulati2020conformer}
A.~Gulati, J.~Qin, C.-C. Chiu, N.~Parmar, Y.~Zhang, J.~Yu, W.~Han, S.~Wang,
  Z.~Zhang, Y.~Wu, and R.~Pang, ``Conformer: Convolution-augmented transformer
  for speech recognition,'' in \emph{Proc. Interspeech 2020}, 2020, pp.
  5036--5040.

\bibitem{hernandez2018ted}
F.~Hernandez, V.~Nguyen, S.~Ghannay, N.~Tomashenko, and Y.~Est{\`e}ve,
  ``{TED-LIUM 3}: Twice as much data and corpus repartition for experiments on
  speaker adaptation,'' in \emph{Proc. International Conference on Speech and
  Computer}.\hskip 1em plus 0.5em minus 0.4em\relax Springer, 2018, pp.
  198--208.

\bibitem{peoplespeech}
``People's {S}peech,'' https://mlcommons.org/en/peoples-speech/, accessed April
  1, 2021.

\bibitem{oneill21}
P.~K. O'Neill, V.~Lavrukhin, S.~Majumdar, V.~Noroozi, Y.~Zhang, O.~Kuchaiev,
  J.~Balam, Y.~Dovzhenko, K.~Freyberg, M.~D. Shulman, B.~Ginsburg, S.~Watanabe,
  and G.~Kucsko, ``{SPGISpeech}: 5,000 hours of transcribed financial audio for
  fully formattedend-to-end speech recognition,'' in \emph{Submitted to
  Interspeech}, 2021.

\bibitem{povey2011kaldi}
D.~Povey, A.~Ghoshal, G.~Boulianne, L.~Burget, O.~Glembek, N.~Goel,
  M.~Hannemann, P.~Motlicek, Y.~Qian, P.~Schwarz \emph{et~al.}, ``{The Kaldi
  speech recognition toolkit},'' in \emph{Proc. 2011 IEEE Workshop on Automatic
  Speech Recognition and Understanding (ASRU)}.\hskip 1em plus 0.5em minus
  0.4em\relax IEEE, 2011.

\bibitem{manohar2017jhu}
V.~Manohar, D.~Povey, and S.~Khudanpur, ``{JHU Kaldi system for Arabic MGB-3
  ASR challenge using diarization, audio-transcript alignment and transfer
  learning},'' in \emph{Proc. 2017 IEEE Automatic Speech Recognition and
  Understanding Workshop (ASRU)}.\hskip 1em plus 0.5em minus 0.4em\relax IEEE,
  2017, pp. 346--352.

\bibitem{pearson1991searching}
W.~R. Pearson, ``Searching protein sequence libraries: Comparison of the
  sensitivity and selectivity of the {Smith-Waterman} and {FASTA} algorithms,''
  \emph{Genomics}, vol.~11, no.~3, pp. 635--650, 1991.

\bibitem{Watanabe2018ESPnet}
S.~Watanabe, T.~Hori, S.~Karita, T.~Hayashi, J.~Nishitoba, Y.~Unno, N.~E.~Y.
  Soplin, J.~Heymann, M.~Wiesner, and N.~Chen, ``{ESPnet}: End-to-end speech
  processing toolkit,'' in \emph{Proc. Interspeech 2018}.\hskip 1em plus 0.5em
  minus 0.4em\relax ISCA, 2018.

\bibitem{Weng2020}
C.~Weng, C.~Yu, J.~Cui, C.~Zhang, and D.~Yu, ``Minimum bayes risk training of
  {RNN}-transducer for end-to-end speech recognition,'' in \emph{Proc.
  Interspeech 2020}.\hskip 1em plus 0.5em minus 0.4em\relax ISCA, 2020, pp.
  966--970.

\bibitem{karita2019comparative}
S.~Karita, N.~Chen, T.~Hayashi, T.~Hori, H.~Inaguma, Z.~Jiang, M.~Someki,
  N.~E.~Y. Soplin, R.~Yamamoto, X.~Wang \emph{et~al.}, ``A comparative study on
  transformer vs rnn in speech applications,'' in \emph{Proc. 2019 IEEE
  Automatic Speech Recognition and Understanding Workshop (ASRU)}.\hskip 1em
  plus 0.5em minus 0.4em\relax IEEE, 2019, pp. 449--456.

\bibitem{vaswani2017attention}
A.~Vaswani, N.~Shazeer, N.~Parmar, J.~Uszkoreit, L.~Jones, A.~N. Gomez,
  L.~Kaiser, and I.~Polosukhin, ``Attention is all you need,'' in
  \emph{{Advances in Neural Information Processing Systems 30 (NIPS 2017)}},
  {2017}.

\bibitem{guo2020recent}
P.~Guo, F.~Boyer, X.~Chang, T.~Hayashi, Y.~Higuchi, H.~Inaguma, N.~Kamo, C.~Li,
  D.~Garcia-Romero, J.~Shi \emph{et~al.}, ``Recent developments on espnet
  toolkit boosted by conformer,'' \emph{arXiv preprint arXiv:2010.13956}, 2020.

\bibitem{kudo2018sentencepiece}
T.~Kudo and J.~Richardson, ``Sentencepiece: A simple and language independent
  subword tokenizer and detokenizer for neural text processing,'' \emph{arXiv
  preprint arXiv:1808.06226}, 2018.

\end{thebibliography}

\end{document}